\documentclass[ reprint,superscriptaddress,aps,pra]{revtex4-1}%
\usepackage[utf8]{inputenc}
\usepackage[english]{babel}
\usepackage{aeguill}
\usepackage{ulem}
\usepackage[justification=justified]{caption}
\usepackage{bbold}
\usepackage{mathptmx}
\usepackage{psfrag,graphicx}
\usepackage{dcolumn}
\usepackage{amsmath,amssymb}
\usepackage{bm}
\usepackage{color}
\usepackage{latexsym}
\usepackage{epstopdf}
\usepackage{color}
\usepackage[english]{babel}
\usepackage{latexsym}
\usepackage{psfrag,graphicx}
\usepackage{amsmath}
\usepackage{amssymb}
\usepackage{amsfonts}
\usepackage{bm}
\usepackage{epstopdf}
\usepackage{appendix}
\usepackage{subfig}
\usepackage[colorlinks=true,citecolor=myblue,linkcolor=myred]{hyperref}
\usepackage{graphicx}%
\providecommand{\U}[1]{\protect\rule{.1in}{.1in}}
\definecolor{mygrey}{gray}{0.35}
\definecolor{myblue}{rgb}{0.2,0.2,0.8}
\definecolor{myzard}{cmyk}{0,0,0.05,0}
\definecolor{mywhite}{rgb}{1,1,1}
\definecolor{mywhite}{rgb}{1,1,1}
\definecolor{myred}{rgb}{1,0.,0.3}
\def\ba{\begin{align}}
\def\enda{\end{align}}
\def\bi{\begin{itemize}}
\def\ei{\end{itemize}}

\def\be{\begin{equation}}
\def\ee{\end{equation}}
\def\bea{\begin{eqnarray}}
\def\eea{\end{eqnarray}}
\def\bse{\begin{subequations}}
\def\ese{\end{subequations}}
\newcommand{\tr}{\mbox{$\mathrm{Tr}$}}
\newcommand{\ket}[1]{|{#1}\rangle}
\newcommand{\bra}[1]{\langle {#1}|}
\newcommand{\average}[1]{\langle {#1} \rangle}

\newcommand{\ketbra}[2]{\left\vert#1\right\rangle\left\langle#2\right\vert}

\newcommand{\Ignore}[1]{ }

\def\i{\text{i}}

\DeclareMathOperator{\arcsinh}{arcsinh}

\begin{document}

\preprint{APS/123-QED}

\title{Analytically solvable $2\times2$ $PT$-symmetry dynamics from su(1,1)-symmetry problems}

\author{R. Grimaudo}
\address{Dipartimento di Fisica e Chimica dell'Universit\`a di Palermo, Via Archirafi, 36, I-90123 Palermo, Italy}
\address{INFN, Sezione di Catania, I-95123 Catania, Italy}
\author{A. S. M. de Castro}
\address{Universidade Estadual de Ponta Grossa, Departamento de F\'{\i}sica, CEP 84030-900, Ponta Grossa, PR, Brazil}
\author{H. Nakazato}
\address{Department of Physics, Waseda University, Tokyo 169-8555, Japan}
\author{A. Messina}
\address{INFN, Sezione di Catania, I-95123 Catania, Italy}
\address{Dipartimento di Matematica ed Informatica dell'Universit\`a di Palermo, Via Archirafi, 34, I-90123 Palermo, Italy}

\date{\today}
\begin{abstract}
A protocol for explicitly constructing the exact time-evolution operators generated by $2 \times 2$ time-dependent $PT$-symmetry Hamiltonians is reported.
Its mathematical applicability is illustrated with the help of appropriate examples. The physical relevance of the proposed approach within gain-loss system scenarios, like two-coupled wave-guides, is discussed in detail.
\end{abstract}

\maketitle

\section{Introduction}

Several decades ago Feshbach described the dynamics of a quantum physical system coupled with a continuum of states, revealing the potentiality of non-Hermitian matrices in Physics \cite{Feshbach}.
Such an effective approach allowed to bring to light and understand a lot of interesting and peculiar physical features of this kind of open systems \cite{Rotter}.
In this respect, to investigate the mathematical properties of these non-Hermitian matrices shed light on the coupling between the system and the surrounding environment \cite{Rotter}.
Such matrices are now called pseudo-Hermitian matrices \cite{Mostafazadeh}, that is, matrices possessing either an entire real spectrum (called in this case quasi-Hermitian matrices) or complex-conjugated eigenvalues \cite{Mostafazadeh}.

The quasi-Hermitian operators commutating with the operator $PT$ ($P$ and $T$ accomplishing the inversion of spatial and time variable, respectively) have been called  $PT$-symmetry Hamiltonians \cite{Bender}.
Their discovery has paved the way towards a new (non-Hermitian) Quantum Physics \cite{ElGanainy}.
From an experimental point of view, indeed, $PT$-symmetry can be established by controllably and appropriately incorporating gain and loss in the system.
Its physical realizability and related interesting applications appear in several fields: photonics \cite{Gbur}, optics, electronics, microwaves mechanics, acoustics and atomic systems (\cite{ElGanainy} and references therein).

This work deals with physical systems living in a two-dimensional Hilbert space and describable by $2 \times 2$ time-dependent quasi-Hermitian $PT$-symmetry matrices.
Despite its apparent simplicity, the two-level case goes beyond a simple toy model.
A wide variety of problems may be traced back to that of a two-state dynamics effectively simulating the most relevant changes happening in complicated quantum-mechanical systems, e.g. in nuclear magnetic resonance \cite{Abragham}, quantum information processing \cite{NC} and polarization optics \cite{Born}.
In the $PT$-symmetry cases, investigated in this paper, the two-level model describes a general sink-source or gain-loss system.
It proves to be very useful for the comprehension of basic theoretical concepts \cite{Bender1} and of many experimental results, e.g. \cite{Bittner} and \cite{Schindler}.
Other physical systems, e.g. coupled waveguides \cite{Yariv,Ruter,Guo}, are exactly described by a two-dimensional $PT$-symmetry Hamiltonian; in such cases, like in other photonic structures \cite{Longhi}, the dynamics of the physical system is governed by a Schr\"odinger-like equation where the time variable is substituted with the spatial one (the propagation direction).

The matrix representation of a sink-source system may be cast as follows
\begin{equation}\label{pt matr}
\tilde{H}=
\begin{pmatrix}
re^{-i\theta} & \gamma \\
\gamma & re^{i\theta}
\end{pmatrix},
\end{equation}
where $r$ and $\theta$ are real.
The diagonal entry $re^{-i\theta}$ ($re^{i\theta}$) describes the energy time evolution in the sink-source \cite{Bender1}.
The off-diagonal parameter $\gamma$ may be interpreted as the coupling existing between the one-state sink and the one-state source \cite{Bender1}.

It has been highlighted \cite{Bender1} that this two-box model is able to capture in its parameter space the passage from a condition where the exchange of energy between the two boxes takes place enabling the system to reach equilibrium to the opposite physical situation.
The existence of such a radical change has been experimentally realized by parametrically changing the coupling constant $\gamma$ \cite{Bender1,Bittner,Guo,Ruter,Schindler} and may be interpreted as an evidence of the transition from unbroken to broken $PT$-symmetry phase.

In this paper we wish to deal with time-dependent $PT$-symmetry Hamiltonian problems.
To this end we introduce a prescribed time-variation of the coupling constant $\gamma$, turning the Hamiltonian model \eqref{pt matr} into
\begin{equation}\label{pt matr td}
\tilde{H}(t)=
\begin{pmatrix}
re^{-i\theta} & \gamma(t) \\
\gamma(t) & re^{i\theta}
\end{pmatrix}.
\end{equation}
Investigating this kind of problems may find applications whenever one is interested in controlling a gain-loss physical system, representable by the Hamiltonian \eqref{pt matr td}.
To find the time-evolution operator generated by $\tilde{H}(t)$ is, generally speaking, a challenging problem strongly dependent on the off-diagonal time-dependent element.
For this reason we search a general mathematical protocol providing examples of time-dependent coupling parameters leading to the prediction of the exact quantum dynamics of the corresponding physical system.
To this end, in the following, strategically, we construct a tool aimed at solving the general dynamical problem generated by the class of $2 \times 2$ quasi-Hermitian su(1,1) Hamiltonians.
This procedure is immediately exploitable for treating dynamical problems characterized by $PT$-symmetry since they constitute a sub-class of the more general class of bi-dimensional su(1,1) dynamical problems \cite{GdCKM}.

In this respect it is worth noting that an approach aimed at individuating exactly solvable two-dimensional su(2) dynamical problems (generally describing a two-level system subjected to time-dependent external driving forces), has been recently presented \cite{Mess-Nak}.
Such a method allowed to find several remarkable exact solutions for $2 \times 2$ dynamical systems \cite{GdCNM,SNGM} as well as deep connections with mathematical issues \cite{MGMN}.
Moreover, it has proved very useful also for the exact solution of more complex dynamical problems, as interacting qudit systems \cite{GMN,GMIV,GBNM,GLSM}.

On the basis of the result \cite{Mess-Nak} and taking account of the `affinity' existing between the SU(2) and the SU(1,1) symmetry-groups (both are sub-groups of the more general SL(2,$\mathbb{C}$) group), we extend the constructive method, successful for su(2) problems, to the su(1,1) case.
Thus, what is reported in the following possesses a twofold interest.
We identify classes of exactly solvable dynamical problems governed by time-dependent bi-dimensional su(1,1) Hamiltonians, which is physically relevant in its own within the framework of pseudo- and quasi-Hermitian matrices \cite{GdCKM}.
Moreover, through this general procedure, we reach the main goal of this paper, getting analytically solvable $PT$-symmetry dynamics with direct physical meaning thanks to the application to the general sink-source model and the gain-loss wave-guide scenarios.

The paper is organized as follows.
In Sec. \ref{Sec II} the class of $2 \times 2$ su(1,1)-symmetry matrices is introduced as well as the generalized von Neumann-Liouville equation related to statistically interpretable dynamics of gain-loss systems described by non-Hermitian Hamiltonians.
Within this context, a parametric solution of the Schr\"odinger equation and the related solvability condition for the dynamical problem are proposed on the basis of the method reported in Ref. \cite{Mess-Nak} and applied to the class of matrices under scrutiny (Appendix \ref{Res Meth SU(1,1)}).
Useful exact examples in the $PT$-symmetry framework are reported in Sec. \ref{Sec. III} together with physical applications in the general sink-source model and in the waveguide optics scenario.
Finally, conclusive remarks can be found in the last section.

\section{$2 \times 2$ su(1,1) matrices, generalized Von-Neumann - Liouville Equation and exact solution of the related dynamical problem} \label{Sec II}

The lowest-dimensional faithful matrix-representation of the group SU$(1,1)$, indeed, consists in the set of all $2 \times 2$ unit-determinant complex matrices $U$, satisfying the relation $\hat{\sigma}^{z}U^{\dagger}\hat{\sigma}^{z}=U^{-1}$, with ${\hat{\sigma}^{x},\hat{\sigma}^{y},\hat{\sigma}^{z}}$ being the standard Pauli matrices.
It is well known that such a representation is not unitary and that the SU$(1,1)$ generators $\hat{K}$'s obey the following algebraic structure \cite{Klimov}
\begin{equation}
\lbrack\hat{K}^{1},\hat{K}^{2}]=-i\hat{K}^{0},\ [\hat{K}^{1},\hat{K}^{0}]=-i\hat{K}^{2},\ [\hat{K}^{2},\hat{K}^{0}]=i\hat{K}^{1},
\end{equation}
and in a $2 \times 2$ representation they are given by $\hat{K}^{0}={\hat{\sigma}^{z}/{2}}$, $\hat{K}^{1}=-i{\hat
{\sigma}^{y}/{2}}$ and $\hat{K}^{2}=i{\hat{\sigma}^{x}/{2}}$.

A null-trace su(1,1) matrix that depends on a real parameter $t$, given by a linear combination, with real $t$-dependent coefficients, of the three generators $\hat{K}^{0}$, $\hat{K}^{1}$ and $\hat{K}^{2}$ of the su(1,1)
algebra, in terms of Pauli matrices, reads
\begin{equation}
H(t)=\Omega(t)\hat{\sigma}^{z} + i\omega_x(t)\hat{\sigma}^{x} - i\omega_y(t)\hat{\sigma}^{y}.
\end{equation}
In the canonical basis $\{\ket{+},\ket{-}\}$ ($\hat{\sigma}^z\ket{\pm}=\pm\ket{\pm}$), the matrix representation of such an operator has the form
\begin{equation}
H(t)=
\begin{pmatrix}
\Omega(t)        & -\omega(t)\\
\omega^{\ast}(t) & -\Omega(t)
\end{pmatrix}
,\label{GHoperator}
\end{equation}
where $\Omega(t)$ ($\omega(t)=\omega_{y}(t)-i\omega_{x}(t)\equiv|\omega(t)|e^{i\phi_{\omega}(t)}$) is a real (generally complex) function of $t$.
Comparing \eqref{GHoperator} with \eqref{pt matr}, it is easy to see that the subclass of $PT$-symmetric su(1,1) Hamiltonians may be obtained by putting $\phi_\omega=\pm\pi/2$ or equivalently by $\omega_y=0$ (up to a unitary transformation consisting in a rotation of $\pi/2$ with respect to the $\hat{y}$ direction, disregarding the constant term in the Hamiltonian \eqref{pt matr} that has no relevance in the dynamics).
In the su(1,1) case, thus, the coupling parameter $\gamma$ between the sink and the source is played by the parameter $\Omega$.

The su(1,1) matrices are pseudo-Hermitian Hamiltonians \cite{Mostafazadeh}, since there exists one linear hermitian (non-singular) matrix $\eta=\ketbra{+}{+}-\ketbra{-}{-}$ such that $H^{\dagger}(t)=\eta H(t)\eta^{-1}$.
Moreover, in the region of the Hamiltonian parameter space defined by $|\omega|^2<\Omega^2$ the su(1,1) Hamiltonian possesses a real spectrum, implying that in such a parameter region it is quasi-Hermitian \cite{GdCKM}.
The nonunitary evolution operator $U(t)$ generated by the Hamiltonian (\ref{GHoperator}) as a solution of the Schr\"odinger equation $i\dot{U}(t)=H(t)U(t)$ ($\hbar=1$) is an element of SU$(1,1),$ and is parametrized
in terms of Cayley-Klein parameters according to
\begin{equation}\label{Ev Op SU(1,1)}
U(t)=
\begin{pmatrix}
\mathfrak{a}(t) & -\mathfrak{b}(t)\\
-\mathfrak{b}^{\ast}(t) & \mathfrak{a}^{\ast}(t)
\end{pmatrix}.
\end{equation}
The entries $\mathfrak{a}(t)$ and $\mathfrak{b}(t)$ are complex-valued functions and satisfy the relation $\det[U]=\left\vert \mathfrak{a}(t)\right\vert ^{2}-\left\vert \mathfrak{b}(t)\right\vert ^{2}=1$.
This implies that $\left\vert \mathfrak{a}(t)\right\vert $ and $|\mathfrak{b}(t)|$ can be greater than one, unlike what occurs in the SU$(2)$ model \cite{Dattoli}.
Then they do not have a direct probabilistic physical meaning.
The probability interpretation must be reconsidered in the SU$(1,1)$ symmetric model since we are addressing a Hamiltonian model with a non-euclidean metric.
To solve this issue it is useful to consider a new non linear equation of motion proposed for the time evolution of the density matrix of a physical system described by a non-Hermitian Hamiltonians $H(t)=K(t)-i\Gamma(t)$ \cite{Sergi}, namely
\begin{equation}\label{Eq non lin NHH}
\dot{\rho}(t)=-i[K(t),\rho(t)]-\{\Gamma(t),\rho(t)\}+2\rho(t) \text{Tr}\{\rho(t) \Gamma(t)\},
\end{equation}
where $K(t)$ and $\Gamma(t)$ are self-adjoint operators.
The general solution of Eq. \eqref{Eq non lin NHH} can be expressed, in terms of the evolution operator $U$ that satisfies the Schr\"odinger equation $i\dot U(t)=H(t)U(t)$, as
\begin{equation}\label{tracepreserving}
\rho(t)={U(t)\rho(0)U^\dagger(t) \over \text{Tr}\left\{ U(t)\rho(0)U^\dagger(t) \right\}},
\end{equation}
and it possesses `good' physical properties that allow possible physical interpretations of the mathematical results \cite{Sergi,Graefe}.
Such an approach has been used to interpret experimental results reported in \cite{Tripathi}.

\subsection{Parametric Solutions of the su(1,1) Dynamical Problem}

It is important to observe that the dynamical problem described by Eq. \eqref{tracepreserving} can be reduced to the solution of the Schr\"odinger equation $i\dot{U}=HU$, giving rise to a system of linear differential equations which may be put in the form
\begin{align}\label{Syst Diff Eq}
\Omega =i[\mathfrak{\dot{a}}\mathfrak{a}^{\ast}-\mathfrak{\dot{b}}\mathfrak{b}^{\ast}],\quad
\omega =i[-\mathfrak{\dot{a}}\mathfrak{b}+\mathfrak{\dot{b}}\mathfrak{a}].
\end{align}
Hereafter we omit the explicit time-dependences, if not necessary.
Following the approach reported in Ref. \cite{Mess-Nak}, with appropriate changes to the class of $2 \times 2$ su(1,1) matrices (see Appendix \ref{Res Meth SU(1,1)}), the two entries $\mathfrak{a}$ and $\mathfrak{b}$ defining the time evolution matrix $U$ in Eq. \eqref{Ev Op SU(1,1)} may be represented as follows
\begin{subequations}\label{Ex Sol Dyn Prob}
\begin{align}
\mathfrak{a} &  = \cosh\left[  \Lambda_{\mathfrak{\theta}}\right] \exp\left[i\left( {\frac{\phi_{\omega}-\phi_{\omega}(0)}{2}}-\frac{\Theta}{2}-\mathcal{R} \right)\right], \\
\mathfrak{b} &  = -i\sinh\left[  \Lambda_{\mathfrak{\theta}}\right]  \exp\left[i \left( {\frac{\phi_{\omega}+\phi_{\omega}(0)}{2}}-\frac{\Theta}{2}+\mathcal{R} \right)\right],
\end{align}
\end{subequations}
where
\begin{subequations}\label{R and Lambda}
\begin{align}
\Lambda_{\mathfrak{\theta}}=&\int_{0}^{t}\left\vert \omega\right\vert \cos\left[  \Theta\right]~dt', \\
\mathcal{R}=&\int_{0}^{t}\frac{\left\vert \omega \right\vert \sin[\Theta]}{\sinh\left[  2\Lambda_{\mathfrak{\theta}}\right] } ~ dt'. \label{R}
\end{align}
\end{subequations}
$\Theta$ is an arbitrary, real, time-dependent function such that $\Theta(0)=0$ and $\Omega$ and $\omega$ appearing in $H$ are related through the following relation 
\begin{equation}\label{Diff Eq Theta}
\dot{\Theta}+2\left\vert \omega\right\vert \sin[\Theta]\coth\left[  2\Lambda_{\mathfrak{\theta}}\right]=2\Omega+\dot{\phi}_{\omega}.
\end{equation}
Equation \eqref{Diff Eq Theta} practically serves as a recipe in the sense that, choosing at will the function $\Theta$ ($\Theta(0)=0$), it determines $\Omega$ ($\omega$) in terms of $\Theta$ and $\omega$ ($\Omega$) making the dynamical problem \eqref{Syst Diff Eq} exactly solvable.

We underline that Eq. \eqref{Diff Eq Theta} provides a possible time variation of the parameter $\gamma(t)$ in Eq. \eqref{pt matr} so that the dynamical problem may be exactly solved.
Indeed, we recall that, by setting $\phi_\omega(t)=\pm\pi/2$ in our framework, $\Omega$ and $|\omega|$ play the role of $\gamma$ and $|r\sin\theta|$, respectively (the sign of the Hamiltonian parameter $r\sin\theta$, depending on the value of $\theta$, is taken into account by the appropriate choice of $\phi_\omega$).
Summing up, the constraint $|r\sin\theta|=const$ (meaning $|\omega(t)|=const.$), leads, through $\Theta$, to a $t$-dependent expression of $\gamma$ making solvable the related dynamical problem.

We emphasize also that Eqs. \eqref{Ex Sol Dyn Prob} and \eqref{Diff Eq Theta} are parametric solutions of the Schr\"odinger equation and this means that they are valid and may be reinterpreted also for problems whose dynamics is ruled by a Schr\"odinger-like equation.
In guided wave optics, for example, a space-dependent Schr\"odinger equation appears \cite{Ruter,Guo}; in such a case the space variable represents the propagation direction of the waves in the guides and a space-dependent coupling may be reached by varyng the distance between the guides \cite{Longhi}.
Thus, e.g. for the physical systems studied in Refs. \cite{Ruter,Guo}, we may interpret Eq. \eqref{Diff Eq Theta} as a prescription how to vary over space the coupling between the guides in such a way to have an exactly solvable system of space-dependent differential equations for the amplitudes of the waves propagating in the guides.
Given an initial condition of the amplitudes, the latter may be written at a certain space point in terms of the solutions in Eqs. \eqref{Ex Sol Dyn Prob}, provided that the time variable is appropriately substituted by the spatial one.

\section{Exact $PT$-Symmetry Examples} \label{Sec. III}

In the following examples we consider $PT$-symmetry cases, that is, we set $\phi_\omega=\pi/2$.

\subsection{Example 1}

If we choose such a parameter $\Theta$ that, given $|\omega|$, satisfies 
\begin{equation}
\int_{0}^{t}|\omega|\cos[\Theta] ~ dt'={\frac{1}{2}}\arcsinh\left[  \kappa\right], \quad \kappa=2\int_{0}^{t}|\omega| dt',
\end{equation}
from Eqs. \eqref{Ex Sol Dyn Prob} we get
\begin{subequations}\label{a and b I case}
\begin{align}
\mathfrak{a}=& \sqrt{{\frac{\sqrt{1+\kappa^{2}}+1}{2}}} \exp\left[-i\left( \frac{\Theta}{2}+\mathcal{R} \right)\right],\\
\mathfrak{b} =& \sqrt{{\frac{\sqrt{1+\kappa^{2}}-1}{2}}} \exp\left[-i \left( \frac{\Theta}{2}-\mathcal{R} \right)\right], \label{b I case}
\end{align}
\end{subequations}
where
\begin{align}\label{Theta and R}
\mathcal{R}={\arcsinh[\kappa] \over 2}.
\end{align}
In this instance, we actually have $\Theta=\arctan \left[ \kappa \right]$ and the relation between the Hamiltonian parameters reads
\begin{equation}\label{Rel 1}
\Omega={|\omega| \over 2} \left[  2+{\frac{1}{1+\kappa^{2}}} \right].
\end{equation}

\begin{figure}[htp]
\begin{center}
\subfloat[][]{\includegraphics[width=0.22\textwidth]{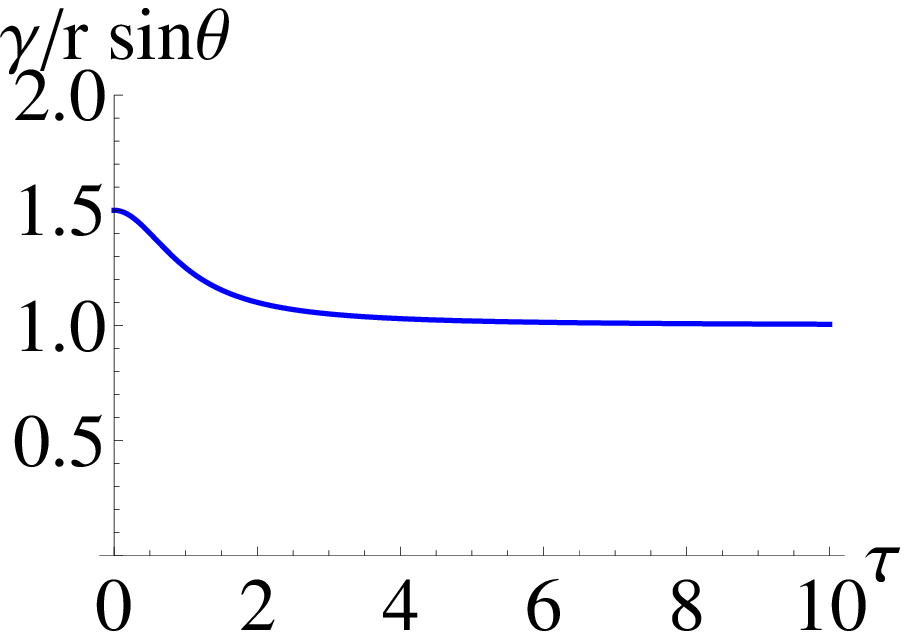}\label{fig:gamma}}
\qquad
\subfloat[][]{\includegraphics[width=0.22\textwidth]{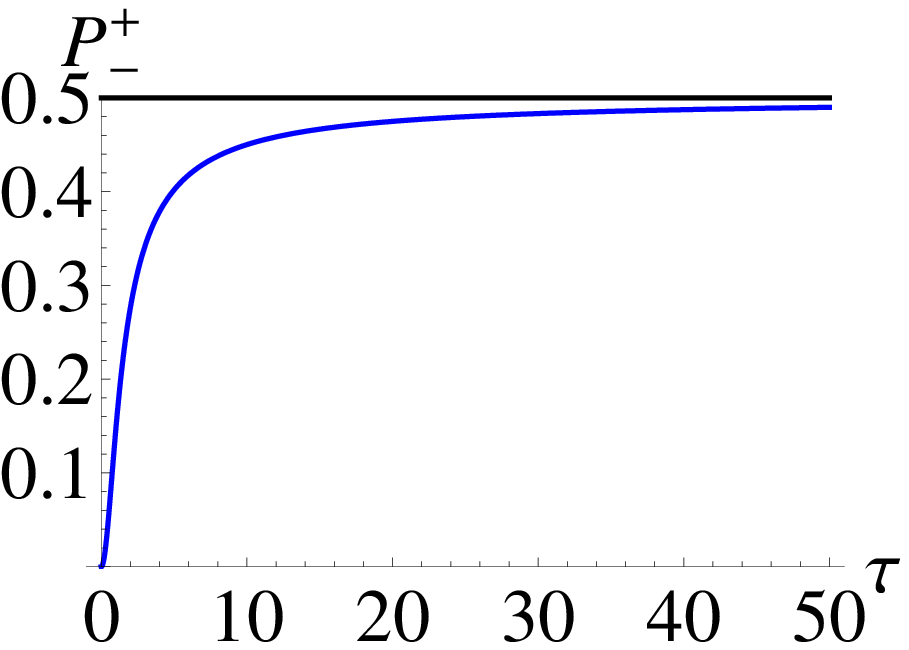}\label{fig:P+-ni}}
\captionsetup{justification=raggedright,format=plain,skip=4pt}%
\caption{(Color online) a) Time evolution of the coupling parameter $\gamma$ in Eq. \eqref{Rel gamma} as a function of $\tau=r \sin(\theta)~t$; b) Time dependence of the transition probability $P_-^+$ in Eq. \eqref{Prob} (for $|\mathfrak{b}|$ in Eq. \eqref{b I case}) as a function of $\tau=r \sin(\theta)~t$. The upper line corresponds to $P_-^+=1/2$.}
\end{center}
\end{figure}

In the framework of gain-loss models we may interpret this relation as a prescription how to vary over time the coupling parameter $\gamma$ in Eq. \eqref{pt matr} between the sink and the source in such a way that the dynamical problem is solved by Eqs. \eqref{a and b I case}.
Supposing $|r\sin\theta|=const.$ (in this case $0<\theta<\pi/2$, in view of the choice $\phi_\omega=\pi/2$; for $\pi/2<\theta<\pi$ we have to set $\phi_\omega=-\pi/2$), we may write Eq. \eqref{Rel 1} as
\begin{equation}\label{Rel gamma}
\gamma(\tau)={|r \sin\theta| \over 2} \left[  2+{\frac{1}{1+\tau^{2}}} \right],
\end{equation}
with $\tau= r \sin(\theta) ~ t$.
The parameter $\gamma$ is plotted in Fig. \ref{fig:gamma} as a function of the dimensionless parameter $\tau$.

If we suppose our system initially prepared in the state $\rho_0=\ketbra{-}{-}$, the probability of finding it in the opposite state, according to Eq. \eqref{tracepreserving}, is
\begin{equation}\label{Prob}
P_-^+=\rho_{11}={|\mathfrak{b}|^2 \over 1+2|\mathfrak{b}|^2},
\end{equation}
where $\rho_{11}$ is the (1,1)-element of the matrix $\rho=U\rho_0U^\dagger/\tr\{ U\rho_0U^\dagger \}$, solution of the equation \eqref{Eq non lin NHH}.
The plot of $P_-^+$ is reported in Fig. \ref{fig:P+-ni} against the dimensionless time $\tau$.
We note that the value reached asymptotically by the transition probability is 1/2.
By Eq. \eqref{Prob} we see, indeed, that 1/2 is the maximum value reachable by the transition probability, precisely when $|\mathfrak{b}| \gg 1$.

A direct physical application of current interest my be found in the framework of coupled wave guides.
In Ref. \cite{Ruter}, for example, a $PT$-symmetry physical scenario of optical beam propagation is investigated.
Under appropriate conditions, the optical-field dynamics of the two coupled wave-guides is described by the following space-dependent Schr\"odinger-like equation
\begin{equation} \label{WG DYN}
i \frac{d}{dz} \begin{pmatrix} E_1 \\ E_2 \end{pmatrix} = 
\begin{pmatrix}
i\epsilon & -\gamma \\
-\gamma & -i\epsilon
\end{pmatrix}
\begin{pmatrix} E_1 \\ E_2 \end{pmatrix},
\end{equation}
where $E_{1,2}$ are the field amplitudes in the first and the second guide, respectively, $\epsilon$ is the effective gain coefficient, $\gamma$ is the coupling constant and $z$ represents the one-dimensional location of the signals in the guides.
In our case, we suppose a spatial configuration of the wave-guides such that the coupling parameter exhibits a spatial-dependence $\gamma \equiv \gamma(z)$.
It is easy to see that, if we apply a unitary transformation accomplishing $\hat{\sigma}_z \rightarrow \hat{\sigma}_x$, $\hat{\sigma}_x \rightarrow -\hat{\sigma}_z$ and $\hat{\sigma}_y \rightarrow \hat{\sigma}_y$, the $PT$-symmetry `Hamiltonian' governing the dynamics of the optical system becomes of the form \eqref{GHoperator}.
In this instance, then, Eq. \eqref{Rel 1}, reading
\begin{equation}
\gamma(\epsilon,z)={\epsilon \over 2} \left[  2+{\frac{1}{1+(\epsilon~z)^{2}}} \right],
\end{equation}
gives us a prescription how to spatially vary the coupling parameter between the two guides, in terms of the constant gain parameter, so that the system \eqref{WG DYN} may be analytically solved.
In such a case the probability $P_+^-$ represents the possibility to transfer the signal in the second wave-guide when it is initially injected in the first one.
The space-dependence of the coupling parameter we are discussing has, then, the effect of asymptotically transferring half part of the initial optical signal from the first channel to the second one.

\subsection{Example 2}

\begin{figure}[htp!]
\begin{center}
\subfloat[][]{\includegraphics[width=0.22\textwidth]{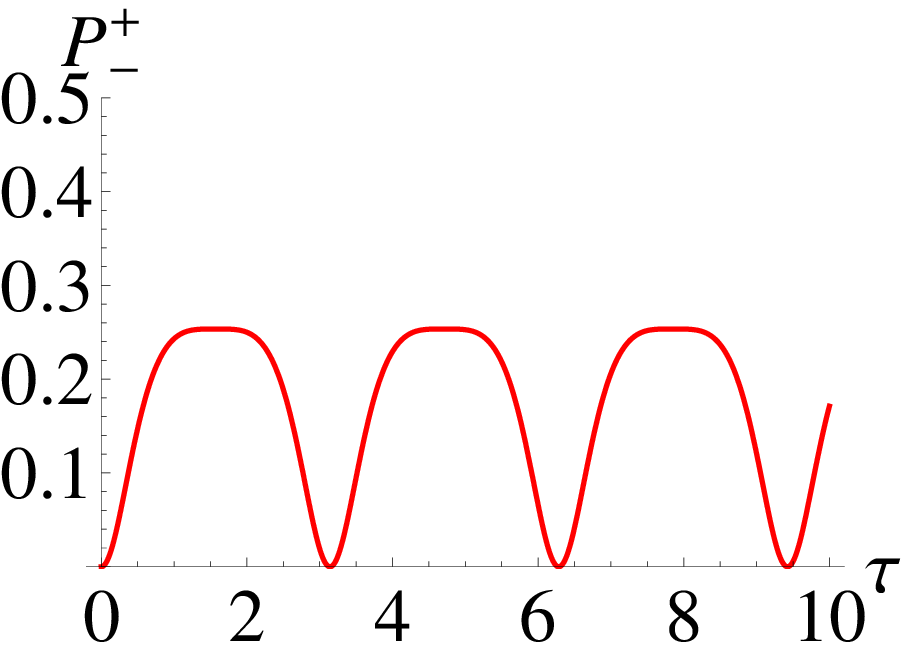}\label{fig:Pnew}}
\qquad
\subfloat[][]{\includegraphics[width=0.22\textwidth]{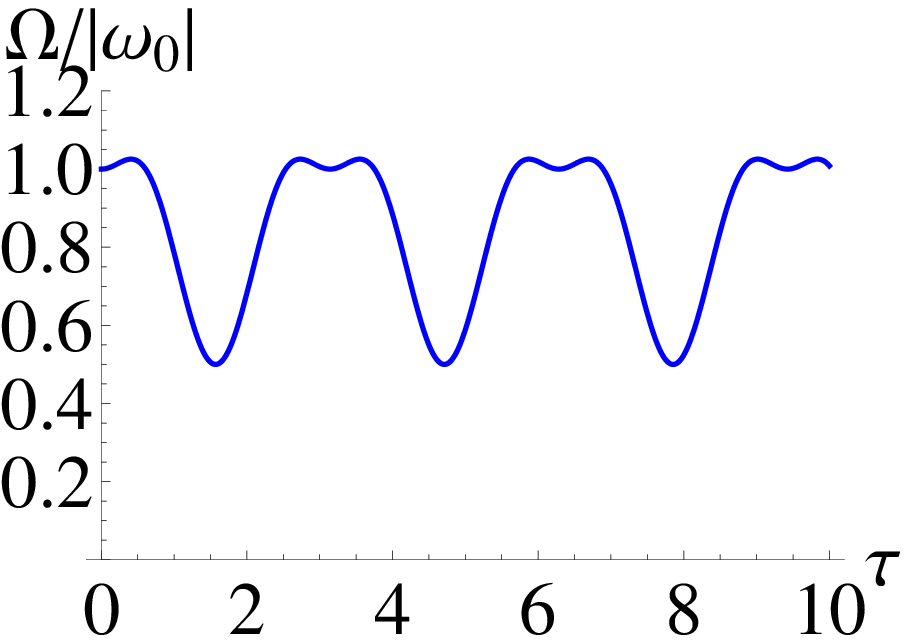}\label{fig:Field}}
\captionsetup{justification=raggedright,format=plain,skip=4pt}%
\caption{(Color online) Time dependence of a) the transition probability $P_-^+$ and b) the parameter $\Omega$ against the dimensionless time $\tau=|\omega_0|t$ for $|\omega|=|\omega_0|\cos^2(\tau)$, $\Theta=\tau$ and $\dot{\phi}_\omega=0$.}
\end{center}
\end{figure}

By putting instead
\begin{equation}
\int_{0}^{t}|\omega|\cos[\Theta] ~ dt'=\arcsinh\left[ \kappa/2 \right]
\end{equation}
it is easy to verify that we get exactly the same expression for $\Theta$ written in Eq. \eqref{Theta and R}, with $\kappa$ replaced by $\kappa/2$, while $\mathcal{R}$ results
\begin{equation}
\mathcal{R}={\frac{1}{2}}\arctan[\kappa/2].
\end{equation}
In this case we have
\begin{subequations}\label{a and b II case}
\begin{align}
\mathfrak{a} &  = \sqrt{1+{\kappa^{2}\over 4}} \exp\left[-i\left( \frac{\Theta}{2}+\mathcal{R} \right)\right],\\
\mathfrak{b} &  = {\kappa \over 2} \exp\left[-i \left( \frac{\Theta}{2}-\mathcal{R} \right)\right], \label{b II case}
\end{align}
and the necessary condition that the Hamiltonian parameters must satisfy results
\end{subequations}
\begin{equation}\label{Rel 2}
\Omega=|\omega|.
\end{equation}
We see that this second choice turns out to be a trivial case.
However, two interesting observations may be developed.
Firstly, it is worth noticing that, developing the same calculation within the framework of su(1,1) matrices, Eq. \eqref{Rel 2} becomes $2\Omega+\dot{\phi}_{\omega }=2\left\vert \omega\right\vert$, which, in turn, is a particular case of the more general solvability condition $2\Omega+\dot{\phi}_{\omega }=2\nu\left\vert \omega\right\vert$ found in \cite{GdCKM} for the class of su(1,1) matrices. 
It is possible to show that, within the class of su(1,1) matrices, this more general relation may be derived through our approach as reported in Appendix \ref{App}.
Secondly, the last example shows how a slight change in the choice of the function $\Theta$ might lead us to a significantly different scenario and then to a substantially different dynamics of the physical system.
This fact, then, underlines the potentiality of the method \cite{Mess-Nak}, here proposed for su(1,1) matrices, in identifying exactly solvable scenarios of possible experimental interest.
\\\\
Before closing this section, we emphasize that in the examples discussed before we were able to find the closed form of all the necessary quantities in order to solve the dynamical problem.
However, if we were interested only in specific physical observables, it might happen that we are requested to find the explicit form of only few quantities.
For example, if we were interested in the study of the transition probability $P_-^+$ or in the knowledge of
\begin{subequations}
\begin{align}
\average{\hat{\sigma}^z}&=\text{Tr}\{\rho\hat{\sigma}^z\}=-1/(|\mathfrak{a}|^2+|\mathfrak{b}|^2), \\
\average{\hat{\sigma}^x}&=\sqrt{{\kappa^2 \over 1+\kappa^2}}\cos[\phi_\omega(t)-\Theta(t)-\pi/2].
\end{align}
\end{subequations}
for $\rho(0)=\ket{-}\bra{-}$, it is sufficient to analytically write the expression of $|\mathfrak{a}|$ and $|\mathfrak{b}|$ ($\Theta$ is chosen at will).
In this way, we are not obliged to solve analytically the expression of the integral $\mathcal{R}$ (Eq. \eqref{R}) involved in the exponentials in Eqs. \eqref{b I case} and \eqref{b II case}, which results in some cases very hard to solve.
Thus, this fact means that, concentrating only on specific physical quantities, the choices of $\Theta$ we may perform and then the classes of exactly solvable models we may identify become wider and wider.
For example, if we choose $|\omega|=|\omega_0|\cos^2(\tau)$ and $\Theta=\tau$ with $\tau=|\omega_0|~t$, although we made relatively simple choices, we are not able to find the analytical expression of the integral $\mathcal{R}$ in Eq. \eqref{R}.
However, we may derive the exact form of the transition probability $P_-^+$ in Eq. \eqref{Prob} plotted in Fig. \ref{fig:Pnew} and, through Eq. \eqref{Diff Eq Theta}, we may write the exact time-dependence of $\Omega$ plotted in Fig. \ref{fig:Field} for $\dot{\phi}_\omega=0$.

%
%
\section{Conclusions}

In conclusion, getting inspired by previous results in other contexts \cite{Mess-Nak}, we have developed a protocol through which we found a parametrization for the solutions of the dynamical problem related to two-dimensional time-dependent quasi-Hermitian su(1,1) Hamiltonians.
Such a result turns out to be of physical interest at the light of the fact that $2 \times 2$ $PT$-symmetry Hamiltonians, describing sink-source or gain-loss systems \cite{Bender1}, are a special sub-class of the su(1,1) matrices \cite{GdCKM}.

This fact allows us to interpret transparently the solvability condition [Eq. \eqref{Diff Eq Theta}] of the dynamical problem from a physical point of view.
Such a relation may be read as the prescription how to vary over time the coupling between the sink and the source in order to controllably drive the dynamics of the whole gain-loss system.
Moreover, we brought to light also the relevance of the result in guided wave optics scenarios \cite{Yariv,Ruter,Guo,Longhi}.
In such cases, the dynamical problem is converted in a space-dependent one obeying to a space-dependent Schr\"odinger-like equation.
Thus, our solutions, provided that the time variable is substituted with the spatial one, are still valid, getting prescriptions how to vary the coupling (space distance) between the wave-guides as to have an analytically solvable model.
The applicability and versatility of our method have been showed with two illustrative examples and by deriving in new terms a previous result recently reported in literature \cite{GdCKM}.

Finally, we emphasize that the parametrized solutions of a $2 \times 2$ su(1,1) dynamical problem here reported possess a more general value.
Indeed, as it happens in the su(2) dynamical case, a higher dimensional su(1,1) dynamical problem may be reduced to the $2 \times 2$ one and its solution may be written in terms of the two parameters $\mathfrak{a}$ and $\mathfrak{b}$ defining the Cayley-Klein parametrization of the evolution operator in Eq. \eqref{Ev Op SU(1,1)} \cite{Ellinas}.

\section*{Acknowledgements}

RG acknowledges support by research funds difc 3100050001d08+, University of Palermo.
ASMC acknowledges the Brazilian agency CNPq financial support Grant No. 453835/2014-7.
This work is partly supported by a Waseda University Grant for Special Research Projects (Project number: 2018K-261).

\appendix

\section{}\label{Res Meth SU(1,1)}

In the following we apply the method reported in ref. \cite{Mess-Nak} to the class of $2 \times 2$ su(1,1) matrices.
The Schr\"odinger equation $i\dot{U}=HU$, with $H$ and $U$ defined in Eq. \eqref{GHoperator} and \eqref{Ev Op SU(1,1)}, respectively, gives rise to a system of linear differential equations which may be put in the form
\begin{align}\label{Syst Diff Eq App}
\Omega =i[\mathfrak{\dot{a}}\mathfrak{a}^{\ast}-\mathfrak{\dot{b}}\mathfrak{b}^{\ast}],\quad
\omega =i[-\mathfrak{\dot{a}}\mathfrak{b}+\mathfrak{\dot{b}}\mathfrak{a}].
\end{align}
Let us introduce the following function
\begin{equation}
X=\int_{0}^{t}\frac{\omega}{\mathfrak{a}^{2}}dt'.
\end{equation}
By such a position and by the second equation in \eqref{Syst Diff Eq App}, we may write respectively
\begin{equation}
\omega=\mathfrak{a}^2\dot{X}, \quad \mathfrak{b}=-i\mathfrak{a}X.\label{sol_b}
\end{equation}
Thus, the relation which the two functions $\mathfrak{a}$ and $\mathfrak{b}$ have to satisfy reads $|\mathfrak{a}|^2{\Large [}1-|X|^2{\Large ]}=1$, implying $|X|^2\leq 1$.
In this way, the first equation in \eqref{Syst Diff Eq} becomes a closed integral-differential equation for $\mathfrak{a}$, namely
\begin{equation}
\mathfrak{\dot{a}}=\left( -i\Omega+{\dot{X}X^{\ast} \over 1-|X|^2} \right) \mathfrak{a},
\label{ea}
\end{equation}
which is solved to yield
\begin{align}
\mathfrak{a} &  =\frac{1}{\left[  1-|X|^2\right]  ^{1/2}} ~ \exp\left[  -i\int_{0}^{t}\Omega ~ dt'+i\int_{0}^{t}\frac{\operatorname{Im}[\dot{X}X^{\ast}]}{1-|X|^2}~dt'\right].
\end{align}

Let us consider an arbitrary complex function $X$ in the form
\begin{equation}
X=A\exp[i\phi],\qquad A(0)=0,
\end{equation}
with $\phi$ and $A$ real functions of time.
The latter must satisfy the condition $A^{2} \leq 1$ due to the condition $|X|^2 \leq 1$.
The function $\dot{\phi}(t)$ has to satisfy
\begin{equation}\label{Eq phi dot}
\dot{\phi}=\frac{\dot{A}}{A}\tan\left[  \Theta\right],
\end{equation}
where $\Theta$ is a real function of time $t$, defined by
\begin{equation}
\Theta=\phi_{\omega}+\phi+2\int_{0}^{t}\Omega~dt'-2\int_{0}^{t}\frac{\dot{\phi}}{1-A^{2}}~dt' - 2 \phi(0).
\end{equation}
By Eq. \eqref{Eq phi dot} we derive $\dot{A}^2+\dot{\phi}^2A^2=\dot{A}^2(1+\tan^2[\Theta])=|\omega|^2(1-|X|^2)$, implying
\begin{equation}\label{A(t) and PhiDot}
A=\tanh\left[\Lambda_{\mathfrak{\theta}} \right], \quad
\dot{\phi}=\frac{2\left\vert \omega\right\vert \sin[\Theta]}{\sinh\left[2\Lambda_{\mathfrak{\theta}}\right]  },
\end{equation}
with
\begin{equation}
\Lambda_{\mathfrak{\theta}}=\int_{0}^{t}\left\vert \omega\right\vert \cos\left[  \Theta\right]~dt',
\end{equation}
so that we have to put $\sin[\Theta(0)]=0$, which is assured by assuming $\phi_\omega(0)=\phi(0)$, if we want to keep parameters well behaved.
We see that the function $A$ in Eq. \eqref{A(t) and PhiDot} satisfies the condition $A^2 \leq 1$.
From the equations in \eqref{A(t) and PhiDot} we find that $\Theta$ must satisfy the following integral-differential equation
\begin{equation}
\dot{\Theta}+2\left\vert \omega\right\vert \sin[\Theta]\coth\left[  2\Lambda_{\mathfrak{\theta}}\right]=2\Omega+\dot{\phi}_{\omega}.
\end{equation}
The solutions $\mathfrak{a}$ and $\mathfrak{b}$ may be written as
\begin{subequations}
\begin{align}
\mathfrak{a} &  = \cosh\left[  \Lambda_{\mathfrak{\theta}}\right] \exp\left[i\left( {\frac{\phi_{\omega}-\phi_{\omega}(0)}{2}}-\frac{\Theta}{2}-\mathcal{R} \right)\right], \\
\mathfrak{b} &  = -i\sinh\left[  \Lambda_{\mathfrak{\theta}}\right]  \exp\left[i \left( {\frac{\phi_{\omega}+\phi_{\omega}(0)}{2}}-\frac{\Theta}{2}+\mathcal{R} \right)\right],
\end{align}
\end{subequations}
with
\begin{equation}
\mathcal{R}=\int_{0}^{t}\frac{\left\vert \omega \right\vert \sin[\Theta]}{\sinh\left[  2\Lambda_{\mathfrak{\theta}}\right] } ~ dt',
\end{equation}
and putting $\Theta(0)=0$.

\section{}\label{App}

To recover the solvability condition $\Omega(t)+\dot{\phi}_\omega(t)/2=\nu|\omega(t)|$, let us assume the function $X$ as
\begin{equation}
X_{\nu}=\nu^{-1}\sin\phi_{\nu}\exp\left[ i(\phi_{\nu}+\phi_\omega^0) \right]
,\quad \phi_{\nu}(0)=0,
\end{equation}
with $\phi_\omega^0=\text{const.}$, and then it is necessary that $\sin^{2}\phi_{\nu}(t)<\nu^{2}$.
Under this assumption and according to the general theory, the transverse field has to be put
\begin{align}\label{omega}
\begin{aligned}
\omega =&\frac{\nu\dot{\phi}_\nu}{\nu^{2}-\sin^{2}\phi_\nu} \\
&\times \exp\left\{ -2i\int_{0}^{t}\Omega ~ dt'+2i\int_{0}^{t}\frac{\nu^{2}\dot{\phi}_\nu}{\nu^{2}-\sin^{2}\phi_\nu} ~ dt' +i\phi_\omega^0 \right\}.
\end{aligned}
\end{align}
Assuming $\nu$ and $\dot{\phi}$ positive we may write
\begin{equation}
\left\vert \omega\right\vert =\frac{\nu\dot{\phi}_\nu}{\nu^{2}-\sin^{2}\phi_\nu}
\label{gc2}
\end{equation}
and from Eq. (\ref{omega}) it is possible to derive
\begin{equation}
2\Omega+\dot{\phi}_{\omega }=2\nu\left\vert \omega\right\vert,\label{RGC}
\end{equation}
being nothing but the relation we were looking for, got through an other approach in Ref. \cite{GdCKM} where its physical reason has been brought to light.
This relation is valid for the dynamical regimes $\nu<1$ and $\nu\geq1$ with $\nu\geqslant0$ and the consistency of the procedure leads to the following expression for $\phi_\nu$
\begin{equation}
\phi_\nu=\arctan\left[ {\nu \over \sqrt{\nu^2-1}}\tan\left( \sqrt{\nu^2-1}\int_0^t|\omega| ~ dt' \right) \right].
\end{equation}
For this case the solutions $\mathfrak{a}_{\nu}$ and $\mathfrak{b}_{\nu}$ can be constructed and acquire different expressions depending on the value of $\nu$.
In the regime $\nu>1,$ we get
\begin{subequations}\label{a b ni sin}
\begin{align}
\mathfrak{a}_\nu & = \left[ \cos(\Lambda_\nu)-i{\nu \over {\sqrt{\nu^2-1}}}\sin(\Lambda_\nu) \right] \exp\left\{  i{\phi_\omega - \phi_\omega^0 \over 2} \right\}, \\
\mathfrak{b}_\nu & = \frac{-i\sin(\Lambda_{\nu}) }{\sqrt{\nu^{2}-1}} \exp\left\{ i{\phi_\omega + \phi_\omega^0 \over 2} \right\},
\end{align}
\end{subequations}
while, in the regime $0\leq\nu<1,$ we have
\begin{subequations}\label{a b ni sinh}
\begin{align}
\mathfrak{a}_\nu &= \left[ \cosh(\Lambda_\nu')-i{\nu \over {\sqrt{1-\nu^2}}}\sinh(\Lambda_\nu') \right] \exp\left\{ i{\phi_\omega - \phi_\omega^0 \over 2} \right\}, \\
\mathfrak{b}_\nu &= -i\frac{\sinh\left[  \Lambda_{\nu}'(t)\right]}{\sqrt{1-\nu^{2}}} \exp\left\{ i{\phi_\omega + \phi_\omega^0 \over 2} \right\},\label{b ni sinh}
\end{align}
\end{subequations}
with
\begin{equation}
\Lambda_{\nu}=\sqrt{\nu^{2}-1}\int_{0}^{t}|\omega| ~ dt', \quad \Lambda_{\nu}'=\sqrt{1-\nu^{2}}\int_{0}^{t}|\omega| ~ dt'.
\end{equation}
\begin{figure}[htp]
\begin{center}
{\includegraphics[width=0.22\textwidth]{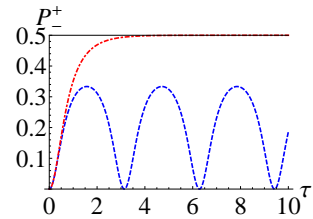}
\captionsetup{justification=raggedright,format=plain,skip=4pt}%
\caption{(Color online) Time dependence of the transition probability $P_-^+$ as a function of $\tau=|\omega_0|t$ for the case \eqref{b ni sinh} with $\nu=1/\sqrt{2}$ (red dot-dashed line) and \eqref{a b ni sin} with $\nu=\sqrt{2}$ (blue dotted line). The upper full line corresponds to $P_-^+=1/2$.}\label{fig:P+-tol}}
\end{center}
\end{figure}
In Fig. \ref{fig:P+-tol} we see the manifestation of the two different regimes for $P_-^+$, namely oscillatory for $\nu > 1$ and asymptotic for $\nu \leq 1$.
It is possible to show \cite{GdCKM} that such different regimes are in general not related to the reality or complexity of the spectrum.
Indeed, in the special case $\Omega=\Omega_0$, $|\omega|=|\omega_0|$ and $\dot{\phi}_\omega=\phi_0$ the Hamiltonian possesses a time-independent spectrum ($E_\pm=\sqrt{\Omega_0^2-|\omega_0|^2}$).
It is possible to show that we may have a $\nu$-based passage from a real to a complex spectrum, keeping the same dynamical regime and, \textit{vice versa}, it may occur a $\nu$-based change of dynamical regime while keeping the reality (or complexity) of the spectrum.
Only for $\phi_0=0$, that is for $PT$-symmetry Hamiltonians and, more ingeneral, for time-independent su(1,1) matrices, we have the coincidence between the two $\nu$-dependent effects, which agree with previous results \cite{Graefe}.
This suggests the fact that the physical feature consisting in the mismatch between the spectrum and the dynamical regime $\nu$-based change, is peculiar of su(1,1) time-dependent pseudo-Hermitian Hamiltonians.


\begin{thebibliography}{99}                                                                                               

\bibitem{Feshbach}
H. Feshbach, Ann. Phys. \textbf{5}, 357 (1958); H. Feshbach, Ann. Phys. \textbf{19}, 287 (1962).

\bibitem{Rotter}
I. Rotter and J. P. Bird, Rep. Prog. Phys. \textbf{78}, 114001, (2015);
I. Rotter, J. Phys. A: Math. Theor. \textbf{42}, 153001, (2009).

\bibitem{Mostafazadeh}
A. Mostafazadeh, J. Math. Phys. \textbf{43}, 205 (2002);
A. Mostafazadeh, J. Math. Phys. \textbf{44}, 974 (2003);
A. Mostafazadeh, J. Phys. A: Math. Gen. \textbf{36}, 7081 (2003);
A. Mostafazadeh and A. Batal, J. Phys. A: Math. Gen. \textbf{37}, 11645  (2004);
A. Mostafazadeh, Phys. Scr. \textbf{82}, 038110 (2010);
A. Mostafazadeh, Int. J. Geom. Methods Mod. Phys. \textbf{07}, 1191 (2010).

\bibitem{Bender}
C. M. Bender and S. Boettcher, Phys. Rev. Lett. \textbf{80}, 5243 (1998);
C. M. Bender, D. C. Brody, and H. F. Jones, Phys. Rev. Lett. \textbf{89}, 270401 (2002);
C. M. Bender, Rep. Prog. Phys. \textbf{70}, 947 (2007).

\bibitem{ElGanainy}
R. El-Ganainy, K. G. Makris, M. Khajavikhan, Z. H. Musslimani, S. Rotter \& D. N. Christodoulides, Nature Physics \textbf{14}, 11–19 (2018).

\bibitem{Gbur}
G. Gbur and K. Makris, Photonics Research \textbf{6}, 5 (2018).

\bibitem{Abragham}
A. Abragham, \textit{The Principles of Nuclear Magnetism} (Clarendon, Oxford, 1961).

\bibitem{NC}
M. A. Nielsen and I. L. Chuang, \textit{Quantum Computation and Quantum Information} (Cambridge University Press, Cambridge, England, 1990).

\bibitem{Born}
M. Born and E. Wolf, \textit{Principles of Optics}, 7th ed. (Cambridge University Press, Cambridge, 1999).

\bibitem{Bender1}
C. M. Bender, B. K. Berntson, D. Parker and E. Samuel, American Journal of Physics \textbf{81}, 173 (2013).

\bibitem{Bittner}
S. Bittner, B. Dietz, U. Gunther, H. L. Harney, M. MiskiOglu, A. Richter, and F. Schafer, Phys. Rev. Lett. \textbf{108}, 024101 (2012).

\bibitem{Schindler}
J. Schindler, A. Li, M. C. Zheng, F. M. Ellis, and T. Kottos, Phys. Rev. A \textbf{84}, 040101(R) (2011).

\bibitem{Yariv}
A. Yariv, IEEE J. Quantum Electron. \textbf{9}, 9 (1973).

\bibitem{Ruter}
C. E. Ruter, K. G. Makris, R. El-Ganainy, D. N. Christodoulides, and D. Kip, Nat. Phys. \textbf{6}, 192 (2010).

\bibitem{Guo}
A. Guo, G. J. Salamo, D. Duchesne, R. Morandotti, M. Volatier-Ravat, V. Aimez, G. A. Siviloglou, and D. N. Christodoulides, Phys. Rev. Lett. \textbf{103}, 093902 (2009).
%

\bibitem{Longhi}
S. Longhi, Laser \& Photon. Rev. \textbf{3}, 243 (2009).

\bibitem{GdCKM}
R. Grimaudo, A. S. M. de Castro, M. K\'us, A. Messina, Phys Rev. A \textbf{98}, 033835 (2018).

\bibitem{Mess-Nak}
A. Messina and H. Nakazato J. Phys. A: Math. Theor. \textbf{47} 445302 (2014).

\bibitem{GdCNM}
R. Grimaudo, A. S. M. de Castro, H. Nakazato and A. Messina, Ann. Phys. (Berlin) 1800198 (2018).

\bibitem{SNGM}
T. Suzuki, H. Nakazato, R. Grimaudo and A. Messina, Sc. Rep. \textbf{8}, 17433 (2018).

\bibitem{MGMN}
L. A. Markovich, R. Grimaudo, A. Messina and H. Nakazato Ann. Phys. (NY) \textbf{385} 522 (2017).

\bibitem{GMN}
R. Grimaudo, A. Messina, H. Nakazato, Phys. Rev. A \textbf{94}, 022108 (2016).

\bibitem{GMIV}
R. Grimaudo, A. Messina, P. A. Ivanov and N. V. Vitanov J. Phys. A \textbf{50} 175301 (2017).

\bibitem{GBNM}
R. Grimaudo,Y. Belousov, H. Nakazato and A. Messina, Ann. Phys. (NY) \textbf{392}, 242 (2017).
    
\bibitem{GLSM}
R. Grimaudo, L. Lamata, E. Solano, A. Messina, Phys. Rev. A \textbf{98}, 042330 (2018).

%

\bibitem{Klimov}
A. B. Klimov, S. M. Chumakov, \textit{A Group Theoretical Approach to Quantum Optics}, 2009 WILEY-VCH Verlag GmbH \& Co. 

\bibitem{Dattoli}
G. Dattoli and A. Torre, J. Math. Phys. \textbf{31}, 236 (1990).

\bibitem{Sergi}
A. Sergi and K. G. Zloshchastiev, Int. J. Mod. Phys. B, \textbf{27}, 1350163 (2013);
A. Sergi, and K. G. Zloshchastiev, Phys. Rev. A \textbf{91}, 062108 (2015);
A. Sergi, and K. G. Zloshchastiev, Journal of Statistical Mechanics: Theory and Experiment 2016.3: 033102 (2016).

\bibitem{Graefe}
D. C. Brody and E. M. Graefe, Phys. Rev. Lett. \textbf{109}, 230405 (2012).

\bibitem{Tripathi}
V. Tripathi, A. Galda, H. Barman, and V. M. Vinokur, Phys. Rev. B \textbf{94}, 041104(R) (2016).

\bibitem{Ellinas}
D. Ellinas, Phys. Rev. A \textbf{45}, 1822 (1992).

\end{thebibliography}
\end{document}